\documentclass{article}
\usepackage{spconf,amsmath,graphicx}
\usepackage{epstopdf}
\usepackage{array}
\usepackage{graphicx}
\usepackage{multirow}
\usepackage{color, soul}
\usepackage{cite}
\usepackage{url}
\usepackage{hyperref}
\usepackage{booktabs}
\usepackage{amsmath}
\usepackage{amssymb}

\title{Explicit intensity control for Accented Text-to-speech}
\name{Rui Liu$^{ 1,*}$\thanks{$*$: Corresponding Author. This research was funded by the High-level Talents Introduction Project of Inner Mongolia University (No. 10000-22311201/002) and the Young Scientists Fund of the National Natural Science Foundation of China (No. 62206136).}, Haolin Zuo$^{ 1}$, De Hu$^{ 1}$, Guanglai Gao$^{ 1}$, Haizhou Li $^{ 2}$}
\address{ $^1$ Inner Mongolia University, China $^2$ The Chinese University of Hong Kong, Shenzhen, China \\
\small{liurui\_imu@163.com, zuohaolin\_0613@163.com, cshood@imu.edu.cn, csggl@imu.edu.cn, haizhouli@cuhk.edu.cn}
}
\begin{document}
%\ninept
%
\maketitle
\begin{abstract}
Accented text-to-speech (TTS) synthesis seeks to generate speech with an accent (L2) as a variant of the standard version (L1). How to control the intensity of accent in the process of TTS is a very interesting research direction, and has attracted more and more attention. Recent work design a speaker-adversarial loss to disentangle the speaker and accent information, and then adjust the loss weight to control the accent intensity. However, such a control method lacks interpretability, and there is no direct correlation between the controlling factor and natural accent intensity. To this end, this paper propose a new intuitive and explicit accent intensity control scheme for accented TTS. Specifically, we first extract the posterior probability, called as ``goodness of pronunciation (GoP)'' from the L1 speech recognition model to quantify the phoneme accent intensity for accented speech, then design a FastSpeech2 based TTS model, named Ai-TTS, to take the accent intensity expression into account during speech generation. Experiments show that the our method outperforms the baseline model in terms of accent rendering and intensity control.
\end{abstract}
\section{Introduction}
\label{sec:intro}

% Unlike conventional neural text-to-speech (TTS) synthesis, which only generates the speaker's standard native speech \cite{wang2017tacotron,shen2018natural,li2019transformerTTS,ren2019fastspeech}, accented TTS synthesis aims to synthesis speech with foreign accent instead of native speech \cite{zhang2019learning}. In other words, 

Accented text-to-speech (TTS) synthesis aims to synthesis speech with foreign accent instead of native speech \cite{zhang2019learning}.
% The majority of the TTS models aim to synthesize high-quality speech of a single speaker/language from the given text and have been extended to support multi speakers/languages.
%Despite the progress, there is an increasing demand for accented L2 speech generation, which requires TTS models to generate high-quality accented L2 speech that well captures the complex and dramatic accent variation information. It is valuable to improve the generation capability of TTS models on various L2 accents.
Note that accent is characterized by a distinctive manner of expression  that is influenced by the mother tongue, social group of speakers, or spoken in a particular region~\cite{loots2011automatic}. 
% Generally, people find it easier to speak with others within their own accent group. 
Therefore, the wide adoption of speech applications, such as chatbot and movie dubbing, calls for the study of accented TTS synthesis \cite{veaux2013voice}. Another important practical application is in the design of tools aimed at improving L2 phonological acquisition in language learners.
% In other words, it requires TTS models to generate high-quality accented L2 speech that well captures the complex and dramatic accent variation information. It is valuable to improve the generation capability of TTS models on various L2 accents \cite{veaux2013voice}.

For accented TTS, some attempts tried to model the accent expression through model interpolation \cite{pucher2010modeling, anumanchipalli2013accent,garcia2014generating,waseem2014speech,kayte2015speech}, variance information prediction \cite{garcia2014generating,waseem2014speech,kayte2015speech},  specific quinphone linguistic features \cite{8462470,liu2020multi} and tone/stress embedding \cite{liu2020multi,shen2018natural}, etc.
However, the accent as perceived in human speech is subtle and at a fine level \cite{munro2006mutual}. How to control the intensity of an accent is still an open challenge \cite{perez2022foreign}. 
% It is difficult to control the intensity of an accent because of the lack of well-defined intensity descriptors.
Wutiwiwatchai et al. \cite{wutiwiwatchai2011accent} proposed an accent level adjustment mechanism for bilingual TTS synthesis, where %, which relies on cross-lingual phone alignment, HMM state mapping and HMM interpolation.
the accent level is adjusted by means of interpolation between HMMs of native phones and HMMs of corresponding foreign phones.
This method provides an effective fine-grained accent intensity control scheme, while it cannot be used in current deep learning TTS models, such as Tacotron \cite{wang2017tacotron,shen2018natural} and FastSpeech \cite{ren2019fastspeech,ren2020fastspeech} based architectures.
In a recent deep learning based multilingual TTS study \cite{zhang2019learning}, the authors employed the domain adversarial training \cite{ganin2016domain} to disentangle the accent identity from the speaker identity where the accent level can be controlled by varying the domain adversarial weight~\cite{zhang2019learning}.
Such an adversarial weight method controls the utterance-level accent intensity of speech by using the model hyper-parameter. There is no direct and measurable correlation between the controlling factor and the natural accent intensity. 
The question is how to characterize the fine-grained phoneme-level accent intensity meaningfully, and employ the intensity to control the synthesis of L2 speech for state-of-the-art TTS models, which is the focus of this paper. 
%
% The early studies of modeling accent expression are carried out on Hidden Markov Models (HMM) \cite{pucher2010modeling, anumanchipalli2013accent,garcia2014generating,waseem2014speech,kayte2015speech}, where we can synthesize accented speech with direct predicting the pitch and engry features of L2 speech \cite{garcia2014generating,waseem2014speech,kayte2015speech} or an intermediate representation between L1 and L2 pronunciation through model interpolation \cite{pucher2010modeling}.   
% Entering the era of deep learning, some works build the accented TTS model with deep neural networks \cite{8462470,liu2020multi}. For example, Henter et al. \cite{8462470} trained a DBLSTM-based acoustic model where the L2 speech is produced by interpolating specific quinphone linguistic features towards phones from the other language that represent non-native mispronunciations. Liu et al.\cite{liu2020multi} built a Tacotron2 \cite{shen2018natural} based multi-lingual TTS model in which the accent expression is manipulated via tone or stress embedding input.

\begin{figure*}[thb!]
    \centering
\centerline{\includegraphics[width=0.92\linewidth]{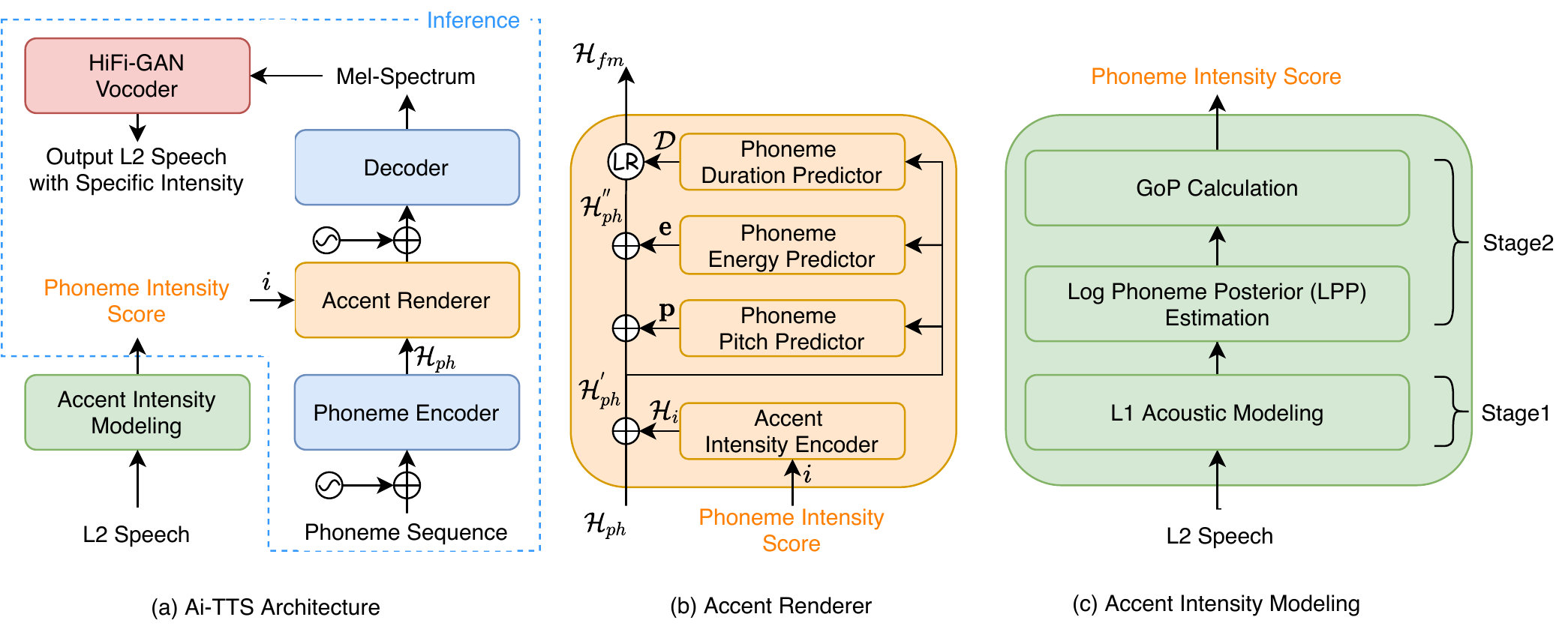}}
    % \vspace{-3mm}
    \vspace{-4mm}
    \caption{The diagrams for the proposed Ai-TTS system. (a) shows the overall model architecture; (b) illustrates the detailed structure of the accent renderer, which includes the accent intensity encoder and the variance adaptor; (c) demonstrates the workflow of the accent intensity modeling.}
    \label{fig:overall}
    \vspace{-5mm}
\end{figure*}

Fortunately, we found that there is a great deal of work in the field of Computer-aided pronunciation training (CAPT) \cite{franco2010eduspeak,yang2022self} to measure the pronunciation of non-native learner. Most of these works assumed that acoustic properties in the learner’s pronunciation are similar to a native English speaker’s acoustics when their pronunciation similarity is high and vice-versa. Considering this, for each phoneme’s in a learner’s utterance, a representative score based on posterior probability of the phoneme models given uttered phoneme speech acoustics, called as \textit{Goodness of Pronunciation} (GoP) \cite{witt2012automatic,witt2000phone} was proposed and achieved remarkable performance.

Inspired by this, in this paper, we proposes a FastSpeech2 based accented TTS model, named \textit{Ai-TTS}, which synthesizes L2 speech by conditioning phoneme-wise accent intensity information. To quantify the fine-grained accent intensity, we utilize a pretrained L1 speech recognition model to calculate the GoP score as the phoneme intensity score for each L2 phoneme. 
During inference, we can control accent expression easily by conditioning intensity labels manually. The experimental results show that our system successfully achieves better accent expressiveness and controllability than the baseline system.

The significant contributions of this work include, 1) We introduce a novel FastSpeech2 based accented TTS synthesis paradigm, named Ai-TTS, that explicitly controls the accent intensity in output speech; 2) We successfully design and implement a fine-grained accent intensity quantization method with accent speech recognition model; 3) We show that the proposed Ai-TTS framework outperforms the baseline models and generates high-quality L2-accented speech.
 
% The rest of this paper is organized as follows. In Section \ref{sec:model}, we propose the Ai-TTS framework. 
% We report the experimental results in Section \ref{sec:exp}. Finally, Section \ref{sec:con} concludes the study.

 \vspace{-3mm}
\section{A\lowercase{i}-TTS: Methodology}
\label{sec:model}
 
 \vspace{-3mm}
\subsection{Model Architecture}
 \vspace{-2mm}
We propose a neural architecture, termed as Ai-TTS, as shown in Fig.\ref{fig:overall} (a) that consists of an accent intensity modeling module, a phoneme encoder, an accent renderer, and a decoder.
The phoneme encoder and decoder are implemented on the basis of FastSpeech2 \cite{ren2020fastspeech}.
The novel accent intensity modeling module aims to learn the phoneme-level accent intensity score for the input L2 speech. Note that the accent intensity modeling module can be treated as the preprocessing operation to label the phoneme intensity score for L2 speech dataset.
The phoneme encoder encodes the input phoneme sequence into phoneme embedding. 
The accent renderer seeks to modulate the input phoneme embeddings with the learned phoneme intensity score and various variance information (including pitch, energy and duration) towards the target accent. Note that the input phoneme intensity score enables all variance information to be affected by the fine-grained accent intensity.
The decoder converts the modulated phoneme embeddings into a mel-spectrum sequence. 
% Note that the accent intensity of the output L2 speech is estimated from the generated mel-spectrum sequence by the additional accent intensity predictor during CAI-TTS training. We impose a consistency constraint loss to minimize the difference between the estimated accent intensity and the expected intensity. 
Finally, the universal HiFi-GAN vocoder \cite{kong2020hifi} is used to synthesize high-quality L2 speech.

% Furthermore, the TTS controls the emotional expression of synthesized speech by controlling intensity with custom labels manually

% In the following subsections, we will introduce the detailed structure of the accent renderer and the workflow for the accent intensity modeling.

 \vspace{-5mm} 
\subsection{Accent Renderer}
 \vspace{-2mm}
Unlike the traditional variance adaptor in \cite{ren2020fastspeech} just adds different variance information such as duration, pitch and energy into the phoneme embeddings, that lacks an accent controlling mechanism, our accent renderer augments the phoneme embeddings with the phoneme accent intensity scalar.
The accent renderer provides phoneme-level accent information according to fine-grained accent intensity. 
As shown in Fig.\ref{fig:overall} (b), the accent renderer consists of 1) an accent intensity encoder, 2) a phoneme pitch predictor, 3) a phoneme energy predictor, and 4) a phoneme duration predictor.

Assume that phoneme embedding $\mathcal{H}_{ph}$ is the phoneme encoder output and the learned phoneme intensity score is $i$.
We implement the accent intensity encoder with a linear layer to transform a real-valued accent intensity score ${i}$ to an intensity embedding vector, $\mathcal{H}_{i}$.
Afterwards, the phoneme-wise intensity embedding $\mathcal{H}_{i}$ is concatenated to the phoneme embedding $\mathcal{H}_{ph}$ to form the accented phoneme embedding $\mathcal{H}^{'}_{ph}$. The phoneme pitch and energy predictors take $\mathcal{H}^{'}_{ph}$ as input and is expected to output more accurate pitch and energy information, that are $\mathbf{p}$ and $\mathbf{e}$ respectively, for L2 speech. We sum the accented phoneme, pitch and energy embeddings to form an augmented accented phoneme embedding $\mathcal{H}^{''}_{ph} = \mathcal{H}^{'}_{ph} + \mathbf{p} +\mathbf{e}$. 
A length regulator (LR) is used to transform the $\mathcal{H}^{''}_{ph}$  to frame-level embeddings  $\mathcal{H}_{fm}$ based on the phoneme duration $\mathcal{D}$ predicted by duration predictor. 

In a nutshell, the accent renderer learns to project the desired accent and its intensity into the input phoneme embedding $\mathcal{H}_{ph}$.
The phoneme-level real-value score $i$ of accent intensity, ranging from 0 to 1, is generated by a novel accent intensity modeling module, which will be described in Sec. \ref{subsec:intensity}.

\subsection{Accent Intensity Modeling}
\label{subsec:intensity}
 \vspace{-2mm}
As mentioned in Sec.\ref{sec:intro},
% when the pronunciation of the L2 learner is similar to the native L1 pronunciation, it indicates that the accent intensity of his pronunciation is weak. On the contrary, when the pronunciation gap between the L2 learner and the native L1 pronunciation is large, it indicates that the accent intensity is strong.
inspired by the CAPT field, we first pretrain the native speech recognition network with the L1 acoustic model, and then quantify the accent intensity for the phoneme sequence of L2 speech by comparing it with the posterior probability between L2 and L1 phonemes. 
% In this subsection, we describe designing L2 accent intensity information via L1 speech recognition modeling. 
As shown in Fig.\ref{fig:overall}(c), the accent intensity modeling is conducted in two stages: 1) Stage1: L1 acoustic modeling stage, and 2) Stage2: accent intensity quantization stage.
% We will introduce them in order.

 \vspace{-3mm}
\subsubsection{Stage1: L1 Acoustic Modeling}
 \vspace{-2mm}
% The traditional GoP score \cite{witt2000phone} is defined under the Gaussian Mixture Model-Hidden Markov Model (GMM-HMM) \cite{bahl1986maximum} or Deep Neural Network-HMM (DNN-HMM) \cite{hu2013new,li2016mispronunciation} based native L1 acoustic model. However, with the development of deep learning, Time-Delay Neural Network (TDNN) based acoustic model \cite{waibel1988consonant} achieves remarkable performance \cite{peddinti2015time,hu2022neural}.

% in terms of speech recognition accuracy. 
% Note that although End-to-End (E2E) speech recognition models \cite{tian2022consistent,zhang2022end} have made great progress recently, the DNN-HMM is more interpretable than E2E models since the GoP calculation needs to involve the calculation of the state transition probability \cite{zhang21x_interspeech}. In addition, DNN-HMM has a more lightweight structure, which can reduce the experimental cycle.
In this work, we employ the Time-Delay Neural Network (TDNN) based acoustic model \cite{waibel1988consonant}
% \footnote{\url{https://github.com/kaldi-asr/kaldi/tree/master/egs/librispeech/s5/local/nnet3}}
, for modelling long term temporal dependencies from short-term acoustic features, with the L1 speech dataset.
The TDNN acoustic model consists of 6 TDNN layers and softmax layer.
The Mel Frequency Cepstral Coefficients (MFCC) and i-vector \cite{5545402} features are extracted as the TDNN input, that are acoustic observations sequences $\mathbf{o}$ as well. 
The initial TDNN layers learn the narrow contexts and the deeper TDNN layers process the hidden activations from a wider temporal context.
As the output, the last softmax layer of TDNN acoustic model can directly output the posterior $\mathcal{P}(\cdot)$ of each phoneme of input speech. More details are referred to \cite{peddinti2015time}.

% Following \cite{sudhakara2019improved}, we train the DNN-HMM acoustic model under the L1 speech dataset.
% The total number of output nodes in DNN is equal to the total number of senones. Each HMM represents a left-to-right three state model. HMM states encode acoustic characteristics of senones. Typically, in the left-to-right HMM, each state is connected to itself by a self-loop transition probability and to the next state by a cross-state transition probability.
 
% In a Context-Dependent DNN-HMM \cite{xx} based acoustic model, the DNN output layer represents the posterior probability, $\mathcal{P}(s|O)$ of each senone, (s) given the uttered acoustic observation sequence O. 

% It is observed that in a DNN-HMM acoustic model, due to state sharing of HMMs, each senone can be associated with many state transition probabilities. Hence, it is non-trivial in DNN-HMM acoustic model to obtain a one-to-one correspondence from senone to state transition probability

After TDNN pretraining, the trained acoustic model takes the phoneme segments of the L2 speech as input to calculate the GoP score, that regarded as the accent intensity score, for each phoneme, which will be described next.

% \begin{table*}[!h]
% \centering
% \small
% \renewcommand\arraystretch{1.2}
% \caption {The comparison of the accent expression for different systems, including expressiveness MOS score, standard deviation ($\sigma$), skewness ($\gamma$), kurtosis ($\mathcal{K}$) and average dynamic time warping (DTW) distances ($\varrho$) for pitch.
% }
 
% \begingroup
% \begin{tabular}{p{4cm}|p{2cm}<{\centering}|p{1.05cm}<{\centering}p{1.05cm}<{\centering}p{1.05cm}<{\centering}p{1.05cm}<{\centering}}
% % \begin{tabular}{l|cccc|c|c}
% \toprule
% $\qquad \qquad  \qquad$System & 
% %
% % \multicolumn{6}{c}{Accent Variance}  \\ \cline{2-7}
%    % & \multicolumn{4}{c|}{Pitch} & Energy & Duration  \\ \cline{2-7}
%    MOS $(\uparrow)$ &$\sigma$ $(\uparrow)$  &  $\gamma$ $(\downarrow)$ &  $\mathcal{K}$ $(\downarrow)$ & $\varrho$ $(\downarrow)$      \\
% \hline 
% Ground Truth  & 4.43 $\pm$ 0.025  & 48.5  & 0.586  & 0.935  & NA  
% \\
%  FastSpeech2 \textcolor{blue}{[1]} & 3.86 $\pm$ 0.031 & 41.3 & 0.632 & 0.992 &  19.22 
% \\ \hline 
% \textbf{Ai-TTS} ($i$=1) \textit{\textbf{(proposed)}} & \textbf{4.01 $\pm$ 0.022} & \textbf{47.2} & \textbf{0.561}  & \textbf{0.944} & \textbf{18.56} \\
% % \hline
% % \quad  w/o accent variance adaptor \& consistency constraint  & xx $\pm$ xx  & xx $\pm$ xx & xx & xx\\
% \bottomrule
% \end{tabular}
 
% \endgroup
% \label{tab:tab2}
% \end{table*}

 \vspace{-4mm}
\subsubsection{Stage2: Accent Intensity Quantization}
 \vspace{-2mm}
To quantify the accent intensity score for all phonemes of L2 speech, the trained TDNN acoustic model takes the L2 speech, instead of L1 speech, as input to extract the posterior for each phoneme $p$.
Afterwards, following \cite{hu2015improved}, the Log Phoneme Posterior (LPP) ratio between the canonical phoneme $p$ and the one phoneme $q$ with the highest score is used to approximate the GoP score:

\vspace{-3mm}
\begin{equation}
   \text{GOP}(p)=\log \frac{\text{LPP}(p)}{\max _{q \in Q} \text{LPP}(q)} 
\end{equation}
\vspace{-3mm}
\begin{equation}
\text{LPP}(p)=\log \mathcal{P}\left(p \mid \mathbf{o} ; t_s, t_e\right)
\end{equation}
where the $Q$ is whole phoneme set. $\mathbf{o}$ is the input acoustic observations. $t_{s}$ and $t_{e}$ are the start and end frame indexes, obtained by forced-alignment, respectively. $\mathcal{P}(p)$ means the prior of phoneme $p$.
Note that the straight way to approximate the $\text{LPP}(p)$ of phoneme segment $p$ is by averaging the frame based posterior $\mathcal{P}(s_{t}|\mathbf{o})$ \cite{hu2015improved}:
 \vspace{-3mm}
\begin{equation}
  \begin{gathered}
\text{LPP}(p) \approx \frac{1}{t_e-t_s+1} \sum_{t=t_s}^{t_e} \log \mathcal{P}\left(p \mid o_t\right) \\
\mathcal{P}\left(p \mid o_t\right)=\sum_{s \in p} \mathcal{P}\left(s \mid o_t\right)
\end{gathered}  
\vspace{-2mm}
\end{equation}
where $s_{t}$ is the senone label of the frame $t$ generated by force alignment with the given canonical phoneme $p$. ${s|s \in p}$ is the states belonging to those triphones whose current phone is $p$.

At last, we follow \cite{liu2022accurate} and normalize the GoP score to [0,1], with 1 as the strongest intensity, as the final accent intensity score $i$ for accent rendering during TTS. 
 
% sudhakara2019improved
 \vspace{-4mm}
\subsection{Run-time Inference}
\vspace{-2mm}
During inference, the Ai-TTS takes the phoneme sequence and synthesizes the controllable L2 speech by conditioning the custom phoneme intensity score manually to achieve explicit intensity control for accented TTS. When all phonemes share a score, it can be viewed as utterance-level control.

\vspace{-3mm}
\section{Experiments and Results}
\label{sec:exp}
\vspace{-2mm}

\subsection{Datasets}
\vspace{-2mm}

\textbf{L1 Speech Dataset:} LibriSpeech corpus \cite{Vassil2020LibriSpeech} is derived from audiobooks that include reading-style speech recorded by 2238 native L1 English speakers, which contains 960 hours of data in the train set. All audios are sampled at 16 kHz and coded in 16 bits. We adopt the ``train\_960\_cleaned'' subset to conduct the TDNN acoustic modeling.

\noindent{\textbf{L2 Speech Dataset:}} We train Ai-TTS on the publicly available L2-ARCTIC corpus \cite{zhao2018l2}, {which includes about 26 hours recordings of accented English from 24 non-native speakers, whose are native in \textit{Hindi, Korean, Mandarin, Spanish, Arabic and Vietnamese}. Two male and two female speakers contributed in each language. } 
In L2-ARCTIC, scripts and their phoneme-level alignment annotations are provided. 
% The scripts for each speaker consist of about 1,130 utterances, resulting in about 27,120 utterances in total. 
% Its phonetic transcription follows the ARPAbet phoneme set \footnote{\url{http://www.speech.cs.cmu.edu/cgi-bin/cmudict}}.
The speech data is sampled at 44.10 kHz and coded in 16 bits. 
For Ai-TTS training, we select the subset of \textit{Mandarin} accent and partition the speech data into training, validation, and test sets at a ratio of 8:1:1.

\vspace{-3mm}
\subsection{Experimental Setup}
\vspace{-2mm}

The phoneme encoder and decoder of Ai-TTS use 6 Feed-Forward Transformer (FFT) blocks. %, which different form the 4 blocks in FastSpeech2 \cite{ren2020fastspeech}.
The dimension of the phoneme embedding $\mathcal{H}_{ph}$ is 256. 
% The phoneme sequence is generated by  the grapheme to phoneme (G2P) conversion toolkit \footnote{https://github.com/Kyubyong/g2p}.
The decoder generates an 80-channel mel-spectrum, which is extracted with 12.5ms frame shift and 50ms frame length, as output.
We downsample all speech files to 22.05 kHz and trimmed leading and trailing silence.
In accent renderer, the accent intensity encoder consists of a linear layer, which encodes the accent intensity score $i$ into a 256 dimensional $\mathcal{H}_{i}$.

We use the Adam optimizer \cite{kingma2014adam} with $\beta_1$ = 0.9, $\beta_2$ = 0.98 and follow the same learning rate schedule in \cite{vaswani2017attention}.
All models are trained with 900k steps to ensure complete convergence.
The codes are written in Python 3.6 using the Pytorch library 1.7.0. The GPU type is NVIDIA Tesla P100 with 24GB GPU memory. 
We employ a pretrained universal HiFI-GAN \cite{kong2020hifi} vocoder for waveform generation.

For TDNN acoustic modeling, we extract the 100 dimensional i-vector and 400 dimensional MFCC features as the TDNN input \footnote{\url{https://github.com/kaldi-asr/kaldi/blob/master/egs/librispeech/s5/run.sh}}. 
The acoustic frame context configuration of TDNN are [-1,0,1], [-1,0,1], [-3,0,3], [-3,0,3], [-3,0,3], [-6,-3,0] in order. 
We follow the Kaldi script \footnote{\url{https://github.com/kaldi-asr/kaldi/blob/master/egs/librispeech/s5/local/nnet3/tuning/run\_tdnn\_1b.sh}} to train TDNN with 128 batch size. The word error rate of the TDNN acoustic model achieved 5.21 \% for various test sets of LibriSpeech on average, which is encouraging. 
As mentioned before, with the help of accent renderer, our Ai-TTS can produce more expressive speech with L2 accent. 
% We validate the Ai-TTS by comparing with \textbf{Ground Truth} L2 speech and the synthesized L2 speech by \textbf{FastSpeech2} \cite{ren2020fastspeech}.
% We validate the Ai-TTS by comparing the L2 speech synthesized by Ai-TTS, FastSpeech2 \cite{ren2020fastspeech} and Ground Truth L2 speech. The subjective and objective evaluations suggest that our Ai-TTS with accent renderer achieves more expressive L2 speech in terms of accent expression. Due to space limit, more details are referred to the accompanied website \footnote{More experimental results and speech samples are available at: \url{https://github.com/ttslr/Ai-TTS}\label{demopage}}.

Note that this paper focuses on the controllability of accent intensity. 
The preliminary objective experiments of accent expression in terms of naturalness and expressiveness are referred at our accompanied website \footnote{More experimental results and speech samples are available at: \url{https://ttslr.github.io/Ai-TTS}\label{demopage}}. The following subsection will investigate the explicit controllability performance for accented TTS.

\vspace{-4mm}
\subsection{Controllability Evaluation on Utterance-level}
\vspace{-2mm}

In this section, we conduct subjective experiments to validate our Ai-TTS
% evaluate the ability of Ai-TTS to adjust the L2 accent intensity explicitly of the synthesized accented speech 
by comparing Ai-TTS with the domain adversarial weight (DAW) control mechanism \cite{zhang2019learning}. Note that DAW is a utterance-level control method, for fair comparison, we set the intensity of all phonemes in Ai-TTS to the same value to achieve utterance-level control.

% Different from the phoneme-level intensity control method of Ai-TTS, DAW uses adversarial weights to control the utterance-level accent intensity. However, if we use one value to define all the phoneme intensity scores in a utterance, it can simulate utterance-level intensity control. 
% To this end, to verify the better interpretability of our explicit intensity control method, we compare the utterance-level intensity control effects of Ai-TTS and DAW for fair comparison.

\begin{figure}[t]
    \centering
    % \vspace{-4mm}
    \centerline{
    \includegraphics[width=0.4\linewidth]{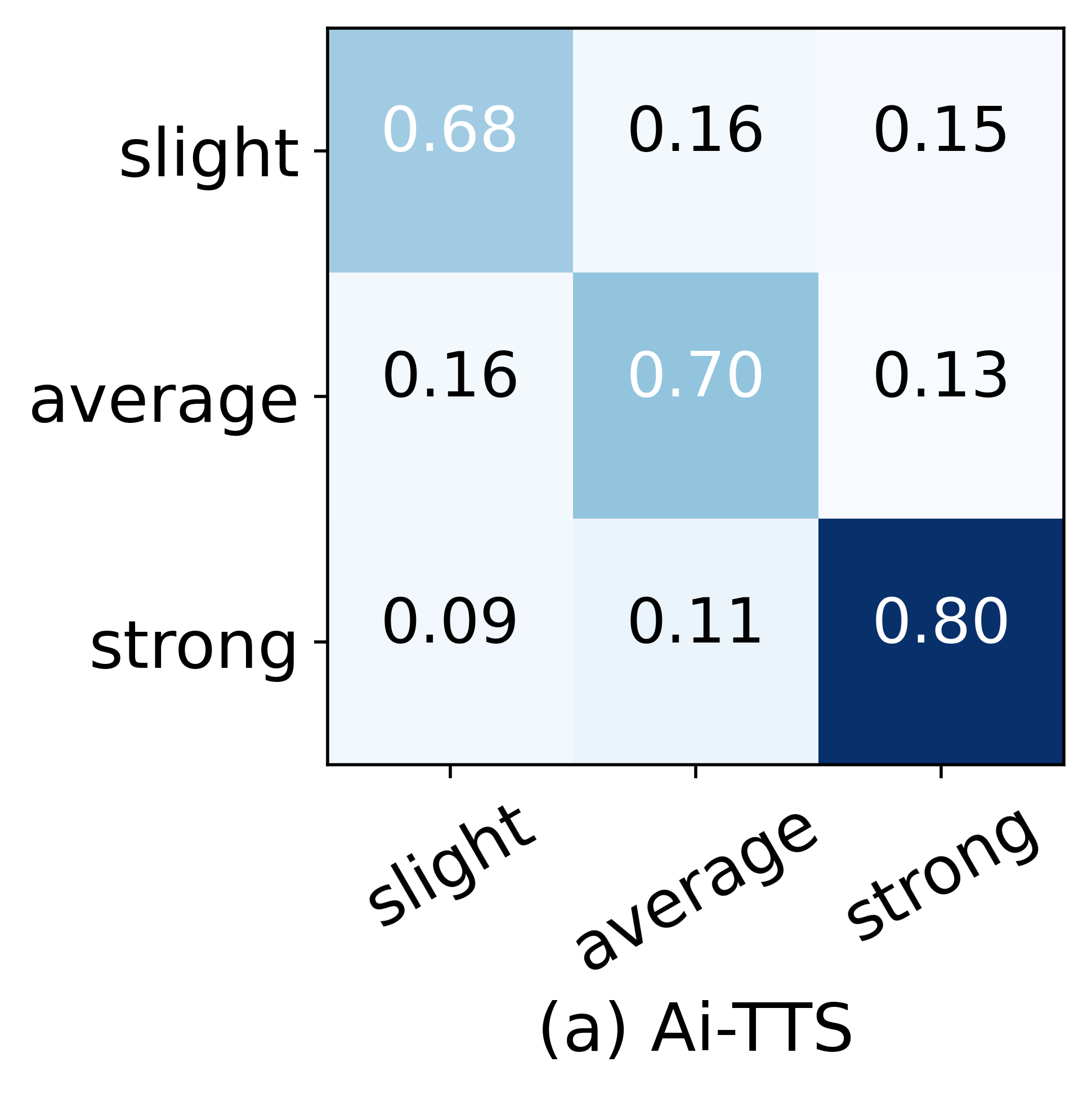}
    \includegraphics[width=0.4\linewidth]{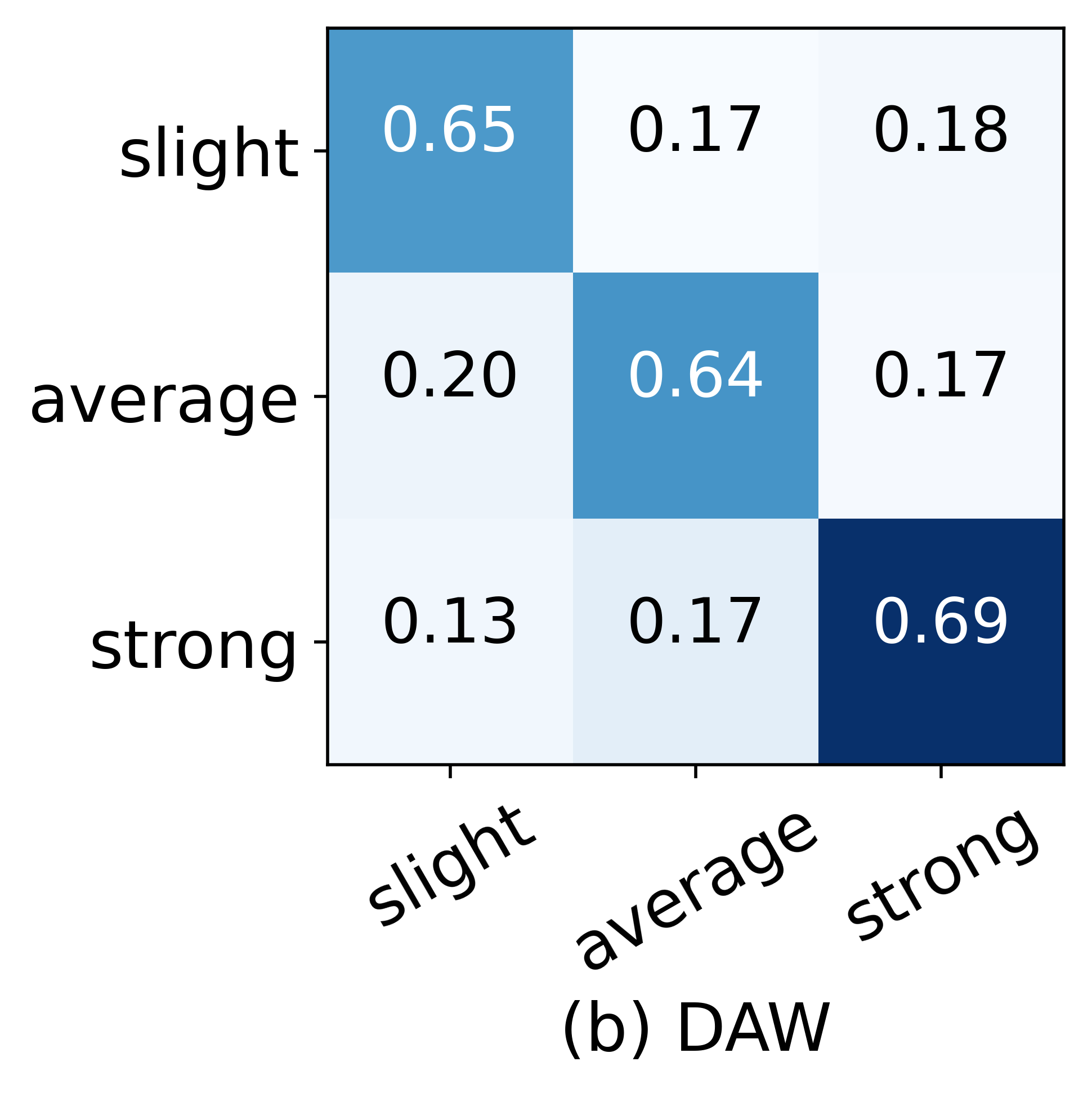}}
    \vspace{-4mm}
    \caption{Confusion matrices between perceived and intended accent intensity categories of synthesized speech.  (a) Ai-TTS;  b) DAW.  The X-axis and Y-axis of the figures represent the perceived and intended category, namely slight, average, and strong.  }
        \vspace{-5mm}
    \label{fig:matirx1}
 
\end{figure}

We first conduct an accent intensity classification experiment. Specifically, for DAW, we follow \cite{zhang2019learning} and set the adversarial weight from 0 to 0.1. We consider the weight value from 0 to 0.03 as `slight', 0.04 to 0.06 as `average' and 0.07 to 0.1 as `strong' in three categories. For Ai-TTS, we consider the intensity scores from 0 to 0.3 as `slight', 0.4 to 0.6 as `average' and 0.7 to 1 as `strong'.
We select 100 utterances from the test set, resulting in 100 samples for both systems. Accordingly, all listeners are instructed to rate the accent intensity category, that are `slight', `average' or `strong', for each sample.
A listener can listen to the samples multiple times when needed.

Fig. \ref{fig:matirx1} reports the intensity confusion matrices. We can find that the Ai-TTS system shows a higher correlation between the perceived and intended accent intensity categories, with a correlation of over 80\%, that is considered a competitive result against other intensity-controlled studies. 
Furthermore, the Ai-TTS system clearly outperforms the contestant. The experiments confirm the superiority of the proposed explicit accent intensity control mechanism.

\begin{figure}[!t]
    \centering
    % \vspace{-21mm}
   \centerline{
    \includegraphics[width=1.1\linewidth]{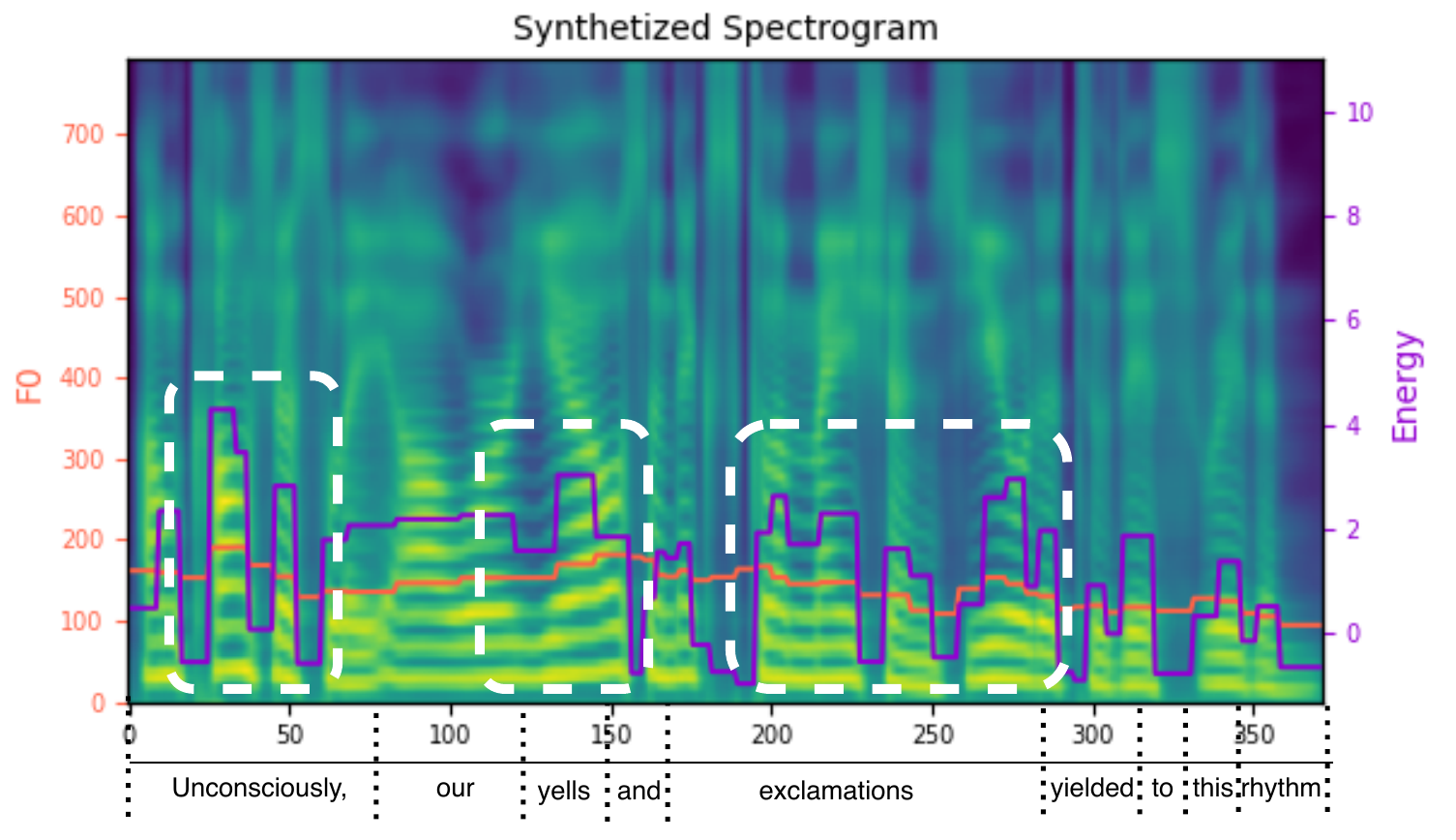}
    }
    % \centerline{
    % \includegraphics[width=0.72\linewidth]{fig/cm-4.png}
    % }
    \vspace{-4mm}
    \caption{Spectrogram, F0 contour and energy of an utterance with specific phoneme-level intensity score. While boxes indicates the acoustic performance with higher accent intensity. 
    % ``\textit{Unconsciously, our yells and exclamations yielded to this rhythm.}''. 
    It is observed that Ai-TTS provides an effective explicit accent intensity control mechanism.}
    \vspace{-4mm}
    \label{fig:matirx2}
 
\end{figure}

\vspace{-4mm}
\subsection{Controllability Evaluation on Phoneme-level}
\vspace{-2mm}

We further evaluate the intensity-controlled speech at phoneme level. Note that F0 (or pitch) and energy are related to accent expression \cite{zhang2019learning}.
Fig. \ref{fig:matirx2} shows an example of spectrogram, F0 contour and energy of an utterance ``Unconsciously, our yells and exclamations yielded to this rhythm.''. Due to space limit, the phoneme sequence ``AH2 N K AA1 N SH AH0 S L IY0 sp AW1 ER0 Y EH1 L Z AE1 N D sp EH2 K S K L AH0 M EY1 SH AH0 N Z sp Y IY1 L D IH0 D T UW1 DH IH1 S R IH1 DH AH0 M'' is omitted. We assign the intensity score of phonemes of words ``Unconsciously'', ``yells'' and ``exclamations'' as 0.9 while those of other phonemes are 0.1. The white boxes in Fig. \ref{fig:matirx2} show that the phonemes with higher accent intensity perform with higher F0 and energy. It indicates that fine-grained accent intensity changes can be easily detected. The Ai-TTS system provides an effective explicit accent intensity control mechanism. We suggest the readers listen to our online demos$^{\ref{demopage}}$.

% \subsubsection{Accent Expression}
% To understand how the accent renderer performs, we randomly select 100 utterances from the test set as the test samples and report the 5-scale Mean Opinion Score (MOS) for three systems, including \textbf{Ground Truth} L2 speech, synthesized L2 speech by \textbf{FastSpeech2} \cite{ren2020fastspeech} and our \textbf{Ai-TTS}.
% For fair comparison, we set $i$ to 5 for all input utterances of Ai-TTS.
% We invite 20 listeners and report the subjective MOS results in the second column of Table \ref{tab:tab2}. It is observed that our Ai-TTS achieves a MOS of 4.01 $\pm$ 0.022, that is significantly higher than \textit{FastSpeech2} baseline and very close to the \textit{Ground Truth}. For objective evaluation, we follow \cite{ren2020fastspeech} and report the moments (including standard deviation ($\sigma$), skewness ($\gamma$) and kurtosis ($\mathcal{K}$)), and average dynamic time warping (DTW) \cite{muller2007dynamic} ($\varrho$) of the pitch distribution between the synthesized L2-accented speech and the ground truth reference in the third to sixth columns of Table \ref{tab:tab2}. It can be seen that the Ai-TTS system is reported with all values that are closer to those of the Ground Truth than FastSpeech2. 

% The subjective and objective evaluations suggest that our Ai-TTS with accent renderer achieves more expressive L2 speech in terms of accent expression.

 \vspace{-5mm}
\section{Conclusion}
\label{sec:con}
\vspace{-2mm}

We have studied a novel TTS model, named Ai-TTS, to control the L2 accent and its intensity explicitly. 
We have conducted a series of experiments on utterance-and phoneme-level intensity control to validate the effectiveness of the Ai-TTS model. The proposed GoP based intensity score outperforms the adversarial weight strategy in terms of interpretability and controllability. This work marks an important step towards controllable rendering of accented TTS synthesis. 
%As our CAI-TTS controls the accent intensity at utterance level, hence in the future, we would like to study fine-grained (eg.  phoneme level) or hierarchical control methods. 
In future work, we plan to further improve the intensity quantitation method.

%\vfill
%\pagebreak
 
% \vfill\pagebreak

% \section{REFERENCES}
% \label{sec:refs}
 
\bibliographystyle{IEEEbib}
{\footnotesize
\bibliography{strings}}

\end{document}